# Mining Frequent Itemsets Using Genetic Algorithm

Soumadip Ghosh[+], Sushanta Biswas[*], Debasree Sarkar[*], Partha Pratim Sarkar[*]

[+]Academy of Technology, Hoogly – 712121,West Bengal, India.
E-mail: soumadip.ghosh@gmail.com

[*]USIC, University of Kalyani, Kalyani, Nadia – 741235, West Bengal, India.
E-mail: mirror@sancharnet.in

**Abstract-** *In general frequent itemsets are generated from large data sets by applying association rule mining algorithms like Apriori, Partition, Pincer-Search, Incremental, Border algorithm etc., which take too much computer time to compute all the frequent itemsets. By using Genetic Algorithm (GA) we can improve the scenario. The major advantage of using GA in the discovery of frequent itemsets is that they perform global search and its time complexity is less compared to other algorithms as the genetic algorithm is based on the greedy approach. The main aim of this paper is to find all the frequent itemsets from given data sets using genetic algorithm.*

**Keywords**- Genetic Algorithm (GA), Association Rule, Frequent itemset, Support, Confidence, Data Mining.

## Introduction

Large amounts of data have been collected routinely in the course of day-to-day management in business, administration, banking, the delivery of social and health services, environmental protection, security and in politics. Such data is primarily used for accounting and for management of the customer base. Typically, management data sets are very large and constantly growing and contain a large number of complex features. While these data sets reflect properties of the managed subjects and relations, and are thus potentially of some use to their owner, they often have relatively low information density. One requires robust, simple and computationally efficient tools to extract information from such data sets. The development and understanding of such tools is the core business of data mining. These tools are based on ideas from computer science, mathematics and statistics. Mining useful information and helpful knowledge from these large databases has thus evolved into an important research area [1, 2].

Data mining has attracted a great deal of attention in the information industry and in society as a whole in recent years, due to the wide availability of huge amounts of data and the imminent need for turning such data into useful information and knowledge. The information and knowledge gained can be used for applications ranging from market analysis, fraud detection, and customer retention, to production control and science exploration.

Frequent pattern mining is an important area of Data mining research. The frequent patterns are patterns (such as itemsets, subsequences, or substructures) that appear in a data set frequently. For example, a set of items, such as milk and bread that appear frequently together in a transaction data set is a *frequent itemset*. A subsequence, such as buying first a PC, then a digital camera, and then a memory card, if it occurs frequently in a shopping history database, is a (*frequent*) *sequential pattern*. A *substructure* can refer to different structural forms, such as subgraphs, subtrees, or sublattices, which may be combined with itemsets or subsequences. If a substructure occurs frequently, it is called a (*frequent*) *structured pattern*. Finding such frequent patterns plays an essential role in mining associations, correlations, and many other interesting relationships among data. Moreover, it helps in data classification, clustering, and other data mining tasks as well.

The process of discovering interesting and unexpected rules from large data sets is known as association rule mining. This refers to a very general model that allows relationships to be found between items of a database. An association rule is an *implication* or *if-then-rule* which is supported by data. The association rules problem was first formulated in [3] and was called the *market-basket* problem. The initial problem was the following: given a set of items and a large collection of sales records, which consist in a transaction date and the items bought in the transaction, the task is to find relationships between the items contained in the different transactions. A typical association rule resulting from such a study could be "90 percent of all customers who buy bread and butter also buy milk" – which reveals a very important information. Therefore this analysis can provide new insights into customer behaviour and can lead to higher profits through better customer relations, customer retention and better product placements. The subsequent paper [4] is also considered as one of the most important contributions to the subject.

Mining of association rules is a field of data mining that has received a lot of attention in recent years. The main association rule mining algorithm, Apriori, not only influenced the association rule mining community, but it affected other data mining fields as well. Apriori and all its variants like Partition, Pincer-Search, Incremental, Border algorithm etc. take too much computer time to compute all the frequent itemsets. The papers [10, 11] contributed a lot in the field of Association Rule Mining (ARM). In this paper, an attempt has been made to compute frequent itemsets by applying genetic algorithm so that the computational complexity can be improved.

## Association Rule Mining (ARM)

Association Rule Mining aims to extract interesting correlations, frequent patterns, associations or casual structures among sets of items in the transaction databases or other data repositories [8]. The major aim of ARM is to find the set of all subsets of items or attributes that frequently occur in many database records or transactions, and additionally, to extract rules on how a subset of items influences the presence of another subset. ARM algorithms discover high-level prediction rules in the form: IF the conditions of the values of the predicting attributes are true, THEN predict values for some goal attributes.

In general, the association rule is an expression of the form X=>Y, where X is antecedent and Y is consequent. Association rule shows how many times Y has occurred if X has already occurred depending on the support and confidence value.

*Support:* It is the probability of item or item sets in the given transactional data base:

support(X) = n(X) / n where n is the total number of transactions in the database and n(X) is the number of transactions that contains the item set X.

Therefore, support (X=>Y) = support(XUY).

*Confidence:* It is conditional probability, for an association rule X=>Y and defined as confidence(X=>Y) = support(XUY) / support(X).

*Frequent itemset:* Let A be a set of items, T be the transaction database and $\sigma$ be the user-specified minimum support. An itemset X in A (i.e., X is a subset of A) is said to be a *frequent itemset* in T with respect to $\sigma$, if support(X)$_T \geq \sigma$.

The problem of mining association rules can be decomposed into two sub-problems:

- Find all sets of items (itemsets) whose support is greater than the user-specified minimum support, $\sigma$. Such itemsets are called *frequent itemsets*.
- Use the frequent itemsets to generate the desired rules. The general idea is that if, say ABCD and AB are frequent itemsets, then we can determine if the rule AB=>CD holds by checking the following inequality

  support({A,B,C,D}) / support({A,B}) $\geq \tau$, where the rule holds with confidence $\tau$.

It is easy to find that the set of frequent sets for a given T, with respect to a given $\sigma$, exhibits some important properties–

- *Downward Closure Property:* Any subset of a frequent set is a frequent set.
- *Upward Closure Property:* Any superset of an infrequent set is an infrequent set.

The above properties lead us to two important definitions–

- *Maximal frequent set:* A frequent set is a *maximal frequent set* if it is a frequent set and no superset of this is a frequent set.

- ▪ *Border set:* An itemset is a *border set* if it is not a frequent set, but all its proper subsets are frequent sets.

Therefore it is evident from the above two dentitions that if we know the set of all maximal frequent sets, we can generate all the frequent sets. Alternatively, if we know the set of border sets and the set of those maximal frequent sets, which are not subsets of any border set, then also we can generate all the frequent sets.

The main association rule mining algorithm, Apriori, also called the *level-wise* algorithm, makes use of the downward closure property. As the name suggests, the algorithm is a bottom-up search, moving upward level-wise in the lattice. This algorithm uses two functions (*candidate generation* and *pruning*) at every iteration. It moves upward in the lattice starting from level 1 till level k, where no candidate set remains after pruning [9].

The algorithm is proposed by R. Agrawal and R. Srikant in 1994 for mining frequent itemsets for Boolean association rules [4]. The name of the algorithm is based on the fact that the algorithm uses *prior knowledge* of frequent itemset properties. The Apriori algorithm pseudo-code for discovering frequent itemsets for mining is given below:

**Pass 1**

1. Generate the candidate itemsets in $C_1$

2. Save the frequent itemsets in $L_1$

**Pass *k***

1. Generate the candidate itemsets in $C_k$ from the frequent itemsets in $L_{k-1}$

   a) Join $L_{k-1}$ $p$ with $L_{k-1}$q, as follows:
      **insert into** $C_k$
      **select** $p$.item$_1$, $p$.item$_2$, . . . , $p$.item$_{k-1}$, $q$.item$_{k-1}$
      **from** $L_{k-1}$ $p$, $L_{k-1}$q
      **where** $p$.item$_1$ = $q$.item$_1$, . . . $p$.item$_{k-2}$ = $q$.item$_{k-2}$, $p$.item$_{k-1}$ < $q$.item$_{k-1}$

   b) Generate all ($k$-1)-subsets from the candidate itemsets in $C_k$

   c) Prune all candidate itemsets from $C_k$ where some ($k$-1)-subset of the candidate itemset is not in the frequent itemset $L_{k-1}$

2. Scan the transaction database to determine the support for each candidate itemset in $C_k$

3. Save the frequent itemsets in $L_k$

Here a *frequent itemset* is an itemset whose support is greater than some user-specified minimum support (denoted $L_k$, where $k$ is the size of the itemset) and a *candidate itemset* is a potentially frequent itemset (denoted $C_k$, where $k$ is the size of the itemset). Apriori and all its variants like Partition, Pincer-Search, Incremental, Border algorithm etc. follow the same functions repeatedly. That is why they take too much computer time to compute all the frequent itemsets.

All the traditional association rule mining algorithms were developed to find positive associations between items. Positive associations refer to associations between items existing in transactions. In addition to the positive associations, negative associations can provide valuable information. In practice there are many situations where negation of products plays a major role. By using Genetic Algorithm (GA) the system can predict the rules which contain negative attributes in the generated rules along with more than one attribute in consequent part. In this regard the contribution of the paper [10] is worth mentioning for finding association rules.

## Genetic Algorithm

Genetic Algorithms (GAs) are adaptive heuristic search algorithm premised on the evolutionary ideas of natural selection and genetic. The basic concept of GAs is designed to simulate processes in natural system necessary for evolution, specifically those that follow the principles first laid down by Charles Darwin of survival of the fittest. As such they represent an intelligent exploitation of a random search within a defined search space to solve a problem.

GAs are one of the best ways to solve a problem for which little is known. They are a very general algorithm and so will work well in any search space. The Genetic Algorithm [5] was developed by John Holland in 1970. GA is stochastic search algorithm modeled on the process of natural selection, which underlines biological evolution [6].

GA has been successfully applied in many search, optimization, and machine learning problems. GA works in an iterative manner by generating new populations of strings from old ones. Every string is the encoded binary, real etc., version of a candidate solution. An evaluation function associates a fitness measure to every string indicating its fitness for the problem [7].

Standard GA apply genetic operators such *selection*, *crossover* and *mutation* on an initially random population in order to compute a whole generation of new strings. GA runs to generate solutions for successive generations. The probability of an individual reproducing is proportional to the goodness of the solution it represents. Hence the quality of the solutions in successive generations improves. The process is terminated when an acceptable or optimum solution is found. GA is appropriate for problems which require optimization, with respect to some computable criterion. The functions of genetic operators are as follows:

1) *Selection:* Selection deals with the probabilistic survival of the fittest, in that, more fit chromosomes are chosen to survive. Where fitness is a comparable measure of how well a chromosome solves the problem at hand.

2) *Crossover:* This operation is performed by selecting a random gene along the length of the chromosomes and swapping all the genes after that point.

3) *Mutation:* Alters the new solutions so as to add stochasticity in the search for better solutions. This is the chance that a bit within a chromosome will be flipped (0 becomes 1, 1 becomes 0).

Essentially, Genetic algorithms are a method of "breeding" computer programs and solutions to optimization or search problems by means of simulated evolution. Processes loosely based on natural selection, crossover, and mutation are repeatedly applied to a population of binary strings which represent potential solutions. Over time, the number of above-average individuals increases and highly-fit building blocks are combined from several fit individuals to find good solutions to the problem at hand.

Not only does GAs provide alternative methods to solving problem, it consistently outperforms other traditional methods in most of the problems link. Many of the real world problems involved finding optimal parameters, which might prove difficult for traditional methods but ideal for GAs.

This generational process is repeated until a termination condition has been reached. Common terminating conditions are:

- A solution is found that satisfies minimum criteria
- Fixed number of generations reached
- Allocated budget (computation time/money) reached
- The highest ranking solution's fitness is reaching or has reached a plateau such that successive iterations no longer produce better results
- Manual inspection
- Combinations of the above

Simple generational genetic algorithm pseudo-code:

```
1. Choose the initial population of individuals
2. Evaluate the fitness of each individual in that population
3. Repeat on this generation until termination: (time limit,
   sufficient fitness achieved, etc.)
   a) Select the best-fit individuals for reproduction
   b) Breed new individuals through crossover and mutation
      operations to give birth to offspring
   c) Evaluate the individual fitness of new individuals
```

```
d) Replace least-fit population with new individuals
```

## Performance Measures

The aim of association rule discovery is the derivation of *if-then-rules* based on the frequent itemsets defined in the previous section. The Apriori algorithm uses a breadth-first search (BFS) approach, first finding all frequent 1-itemsets and then discovering 2-itemsets and continues by finding increasingly larger frequent itemsets. The data is a sequence $x^{(1)}, \ldots, x^{(n)}$ of binary vectors. We can thus represent the data as an $n$ by $d$ binary matrix, where $n$ is the number of data records and $d$ is number of items [11]. At each level k the data base is scanned to determine the support of items in the *candidate itemset* $C_k$. The major determining parameter for the complexity of the Apriori algorithm is:

$$C = \sum_k m_k \text{ where } m_k = |C_k|.$$

We know that $m_1 = d$ as one needs to consider all single items. Furthermore, one would not have any items which alone are not frequent and so one has $m_2 = d(d\text{-}1)/2$. Thus we get the lower bound for $C$:

$$C \leq m_1 + 2m_2 = d^2.$$

As one sees in practice that this is a large portion of the total computations one has a good approximation $C \approx d^2$. Including also the dependence on the data size we get for the time complexity of Apriori:

$$T = O(d^2 n).$$

Thus we have scalability in the data size but *quadratic* dependence on the dimension or number of attributes.

But a genetic algorithm consists of a population and an evolutionary mechanism. The population is a collection of individuals which represent potential solutions through a mapping called a coding. The population is ranked with the help of fitness function. We apply genetic algorithm on the selected population from the database and compute the fitness function after each step.

The paper [12] proposes three different measures for performance evaluation of GAs based on observations in experiments by simulation: The *likelihood of optimality*, the *average fitness value* and the *likelihood of evolution leap*. Based on the above measures, a term has been introduced, *cut-off generation k*, i.e., how many generations a GA should be executed in each run.

Considering $C = kr$ be the total computation cost given to execute the GA, where $r$ is the number of repeated runs; the best cut-off generation is the number $k$ of generations which maximizes the performance with respect to a particular measure. If $C$ is fixed, we want to find $k$ maximizing the term $(1- p(k))^r$, where p(k) denotes the probability that a GA produces an optimal solution within k generations. If the value of $(1- p(k))^r$ is fixed, we want to find k minimizing $C = kr$. Surely this indicates that the GA based simulations can be completed in *linear* time.

Therefore it follows from the above theoretical discussion that GA based solution provides significant improvement in computational complexity in comparison with Apriori algorithm and all its variants.

## Methodology

In this paper the Genetic algorithm (GA) is applied on large data sets to discover the frequent itemsets. We first load the sample of records from the transaction database that fits into memory. The genetic learning starts as follows. An initial population is created consisting of randomly generated transactions. Each transaction can be represented by a string of bits.

Our proposed genetic algorithm based method for finding frequent itemsets repeatedly transforms the population by executing the following steps:

(1) *Fitness Evaluation:* The fitness (i.e., an objective function) is calculated for each individual.

(2) *Selection:* Individuals are chosen from the current population as parents to be involved in recombination.

(3) *Recombination:* New individuals (called offspring) are produced from the parents by applying genetic operators such as crossover and mutation.

(4) *Replacement:* Some of the offspring are replaced with some individuals (usually with their parents).

One cycle of transforming a population is called a generation. In each generation, a fraction of the population is replaced with offspring and its proportion to the entire population is called the generation gap (between 0 and 1).

## Results

We use the example database from A.K. Pujari's book and the different databases from this URL (`http://www2.cs.uregina.ca/~dbd/cs831/notes/itemsets/datasets.php`) to show the effectiveness of the proposed method by using MATLAB software.

Initially we have considered the transaction database (example database from A.K. Pujari's book, *Chapter 4: Association Rules*), T, which contains 15 records listed in Table 1 (see below). Let us consider the set of items, A = {A1, A2, A3, A4, A5, A6, A7, A8, A9} and assume σ = 20%.

Since T contains 15 records, it means that an itemset that is supported by at least 3 transactions is a frequent set. Here presence of 1 at i-th position indicates occurrence of the item[i] in a transaction. Similarly presence of 0 at j-th position indicates absence of item[j].

*Table 1.*

| A1 | A2 | A3 | A4 | A5 | A6 | A7 | A8 | A9 |
|----|----|----|----|----|----|----|----|----|
| 1  | 0  | 0  | 0  | 1  | 1  | 0  | 1  | 0  |
| 0  | 1  | 0  | 1  | 0  | 0  | 0  | 1  | 0  |
| 0  | 0  | 0  | 1  | 1  | 0  | 1  | 0  | 0  |
| 0  | 1  | 1  | 0  | 0  | 0  | 0  | 0  | 0  |
| 0  | 0  | 0  | 0  | 1  | 1  | 1  | 0  | 0  |
| 0  | 1  | 1  | 1  | 0  | 0  | 0  | 0  | 0  |
| 0  | 1  | 0  | 0  | 0  | 1  | 1  | 0  | 1  |
| 0  | 0  | 0  | 0  | 1  | 0  | 0  | 0  | 0  |
| 0  | 0  | 0  | 0  | 0  | 0  | 0  | 1  | 0  |
| 0  | 0  | 1  | 0  | 1  | 0  | 1  | 0  | 0  |
| 0  | 0  | 1  | 0  | 1  | 0  | 1  | 0  | 0  |
| 0  | 0  | 0  | 0  | 1  | 1  | 0  | 1  | 0  |
| 0  | 1  | 0  | 1  | 0  | 1  | 1  | 0  | 0  |
| 1  | 0  | 1  | 0  | 1  | 0  | 1  | 0  | 0  |
| 0  | 1  | 1  | 0  | 0  | 0  | 0  | 0  | 1  |

The initial population was 20 and crossover was chosen randomly. The mutation probability was taken 0.05. The frequent itemsets with user-specified minimum support ($\sigma$) $\geq$ 20% generated for the given database are listed in Table 2 as follows (though the MATLAB output shows the frequent itemsets in a binary string format). Obviously the result matches with the result found from Apriori algorithm and all its variants.

*Table 2.*

| Frequent 1-itemsets: | [2, 3, 4, 5, 6, 7, 8] |
|----------------------|----------------------|
| Frequent 2-itemsets: | [2 3, 2 4, 3 5, 3 7, 5 6, 5 7, 6 7] |
| Frequent 3-itemsets: | [3 5 7] |

As described earlier, the implementation of GAs is also applied on different large data sets taken from the above mentioned URL (for example, *10000* x *8 Database*, *Plant Cell Signaling*, *Zoo*, *Tic Tac Toe*, *Synthetic #1*, and *Synthetic #2* etc.) by using MATLAB software. In every case we got satisfactory results from our experiments. For example, when we tested our proposed GA based method on the huge data set of *10000* x *8 Database* (consisting of 10000 records) we also achieved success which surely proves the effectiveness of the proposed method.

## Conclusion and Future Work

We have dealt with a challenging association rule mining problem of finding frequent itemsets using our proposed GA based method. The method, described here is very simple and efficient one. This is successfully tested for different large data sets. The results reported in this paper are correct and appropriate. However, a more extensive empirical evaluation of the proposed method will be the objective of our future research. We also intend to compare the performance of our GA based method proposed in this paper with the FP-tree algorithm. The incorporation of other interestingness measures mentioned in the literature is also part of our planned future work.